\documentclass[graybox]{svmult}

\usepackage{mathptmx}       
\usepackage{helvet}         
\usepackage{courier}        
\usepackage{type1cm}        
\usepackage{makeidx}         
\usepackage{graphicx}        
\usepackage{multicol}        
\usepackage[bottom]{footmisc}

\usepackage{comment}
\usepackage{subcaption}
\usepackage{mwe}
\usepackage{multirow}
\usepackage[inline]{enumitem}
\usepackage{amsmath}
\usepackage{caption,booktabs}
\usepackage{amssymb}

\makeindex             

\begin{document}

\title*{Deep Agent: \\Studying the Dynamics of Information Spread and Evolution in Social Networks}
\titlerunning{Deep Agent Modeling Framework}
\author{Ivan Garibay$^{1,*}$, Toktam A. Oghaz$^1$, Niloofar Yousefi$^1$, Ece {{\c{C}}}i{{\u{g}}}dem Mutlu$^1$, Madeline Schiappa$^1$, Steven Scheinert$^1$, 
Georgios C. Anagnastopoulos$^2$, Christina Bouwens$^1$, 
Stephen M. Fiore$^1$, Alexander Mantzaris$^1$, John T. Murphy$^3$, William Rand$^4$, Anastasia Salter$^1$, Mel Stanfill$^1$, Gita Sukthankar$^1$, 
Nisha Baral$^1$, Gabriel Fair$^5$,
Chathika Gunaratne$^1$, Neda B. Hajiakhoond$^1$, 
Jasser Jasser$^1$, Chathura Jayalath$^1$, 
Olivia Newton$^1$, Samaneh Saadat$^1$, 
Chathurani Senevirathna$^1$,
Rachel Winter$^1$, and Xi Zhang$^2$}
\authorrunning{Garibay et al.}
\institute{$^1$University of Central Florida 
,$^2$Florida Institute of Technology, $^3$University of Chicago\\ $^4$North Carolina State University, $^5$University of North Carolina at Charlotte \\
$^{*}$Corresponding author: Ivan Garibay
\at \email{igaribay@ucf.edu}
}
%
%
\maketitle

\abstract*{This paper explains the design of a social network analysis framework, developed under DARPA's SocialSim program, 
with novel architecture that models  
human emotional, cognitive and social factors.
Our framework is both theory and data-driven, and utilizes domain expertise. Our simulation effort helps understanding how information flows and evolves in social media platforms. We focused on modeling three information domains: cryptocurrencies, cyber threats, and software vulnerabilities for the three interrelated social environments: GitHub, Reddit, and Twitter. We participated in the SocialSim DARPA Challenge in December 2018, in which our models were subjected to an extensive performance evaluation for accuracy, generalizability, explainability, and experimental power. 
This paper reports the main concepts and models, utilized in our social media modeling effort in developing a multi-resolution simulation at the user, community, population, and content levels.}

\abstract{This paper explains the design of a social network analysis framework, developed under DARPA's SocialSim program, 
with novel architecture that models  
human emotional, cognitive and social factors.
Our framework is both theory and data-driven, and utilizes domain expertise. Our simulation effort helps understanding how information flows and evolves in social media platforms. We focused on modeling three information domains: cryptocurrencies, cyber threats, and software vulnerabilities for the three interrelated social environments: GitHub, Reddit, and Twitter. We participated in the SocialSim DARPA Challenge in December 2018, in which our models were subjected to an extensive performance evaluation for accuracy, generalizability, explainability, and experimental power. 
This paper reports the main concepts and models, utilized in our social media modeling effort in developing a multi-resolution simulation at the user, community, population, and content levels.}

\section{Introduction}
\label{sec:1}
Emerging Online Social Networks (OSNs) have revolutionized the public information environment in an unprecedented way. 
Thus, it is crucial to study the process of the spread and evolution of online information to understand the reach and impact of news, ideas, and knowledge in OSNs. An accurate and scalable computational simulation of this process could potentially help combat misinformation campaigns by adversaries, efficiently deliver critical information to local populations during disaster relief operations, and contribute to social construction and policy designs that rely on information dissemination.

Despite progress in this field of research, current computational approaches to social and behavioral simulations have not been well-positioned to uncover the underlying dynamics that explain the inner workings and reasons for the selection and diffusion of information in online social platforms.
Current approaches to online social dynamic simulations fall into three main categories: 
\begin{enumerate*}[label={\Roman*)}]
\item the statistical analysis and modeling of a particular phenomenon such as ``information evolution'' using a particular dataset and fitting a statistical model to the data, for instance, \cite{adamic2016information}; \item the statistical physics approach using the Agent-Based Model (ABM) simulation as an extension of dynamic equation modeling; and \item approaches using ABMs through ``translating'' a theoretical model into the agent-based framework \cite{rand2015agent}. 
\end{enumerate*}
Although the first approach can be used in econometrics to predict the economic outcomes a few months ahead, it fails to predict 
rich system dynamics (such as during a financial crisis) correctly, and does not take complex human dynamics into account,
though it models a single dataset accurately. 
The second approach is typically based on a ``Brownian agent'' \cite{vsuvakov2012agent}), and applies agent-based models in a very different capacity than the more standard practice of using them as a complex systems modeling tool. The Brownian agents are restricted by the stochastic physics framework in which they are embedded, 
resulting in less capability in capturing complex dynamics behavior. 
Lastly, the third approach focuses on replicating a single phenomenon; the agents mirror a single set of equations that are focused on an observable macro-pattern instead of the deep cognitive mechanisms that drive human behavior. 
As a result of not modeling deep human emotional, cognitive and social factors that determine social behavior outcomes, all three approaches lead to potential over-fitting on a single dimension of data.

Although frameworks like the Agent-Zero \cite{epstein2014agent_zero} and Homo Socialis \cite{gintis2015homo} offer theoretical solutions to modeling the true complexity of human dynamics driven by deep neurocognitive underpinnings that are at the core of any human social activity, including the spread and evolution of information, these deep models are limited to modeling conceptual problems. Furthermore, 
simpler models are preferred for real world problems as they allow for the parameters tuning directly associated with modeling a particular dataset, simpler models 
cannot simultaneously replicate multiple complex phenomena such as various aspects of human dynamics, including information cascading, gatekeeper's identification, information evolution, and persistent minorities. 

To overcome the shortages in modeling the online social platforms, the Defense Advanced Research Projects Agency (DARPA) announced the ``Computational Simulation of Online Social Behavior (SocialSim)'' program to develop innovative technologies for high-fidelity computational simulation of online social behavior. Responding to this DARPA program,
our team proposed and implemented a novel simulation framework
that enables revolutionary advances in simulation of information spread and evolution on social media on a large scale. 
Our team accomplished this by \begin{enumerate*}[label={\Roman*)}]
\item modeling social dynamics using a network of computational agents endowed with deep neurocognitive capabilities, \item creating a family of plausible social dynamic models assembled from modularized sub-components, and \item utilizing machine Learning algorithms and HPC Cloud Computing for model discovery, refinement, and testing.
\end{enumerate*}

This paper explains the design and concepts related to our framework and agent-based models for social network analysis 
using large volumes of data from GitHub, Reddit, and Twitter 
with the aim of 
better understanding of online social behaviors. We proposed the Deep Agent Framework (DAF), which operationalizes social theories of human behavior and social media into optimized generative simulation capabilities. Additionally, we developed Multiplexity-Based Model (MBM) which is an agent-based model designed based on concepts from graph theory, that simulates online social network evolution. 


\section{Challenge Problem Description}
\label{sec:2}

The challenge problem was designed to develop a multi-resolution simulation at the content, user, community, and population levels. Thus, a total of 57 accuracy measurements and metrics were designed to evaluate the participant models. 
The challenge evaluation procedure applied
a combination of various metrics and measurements over four dimensions: accuracy, generalizability, explainability, and experimental power. Table \ref{tab:1} in the Appendix section contains the evaluation metrics and a performance comparison between the agent-based models for the community, content, population, and user level interactions. 




\section{Methodology}
\begin{figure*}
        \centering
        \begin{subfigure}[b]{\textwidth}
            \centering
            \includegraphics[width=\textwidth]{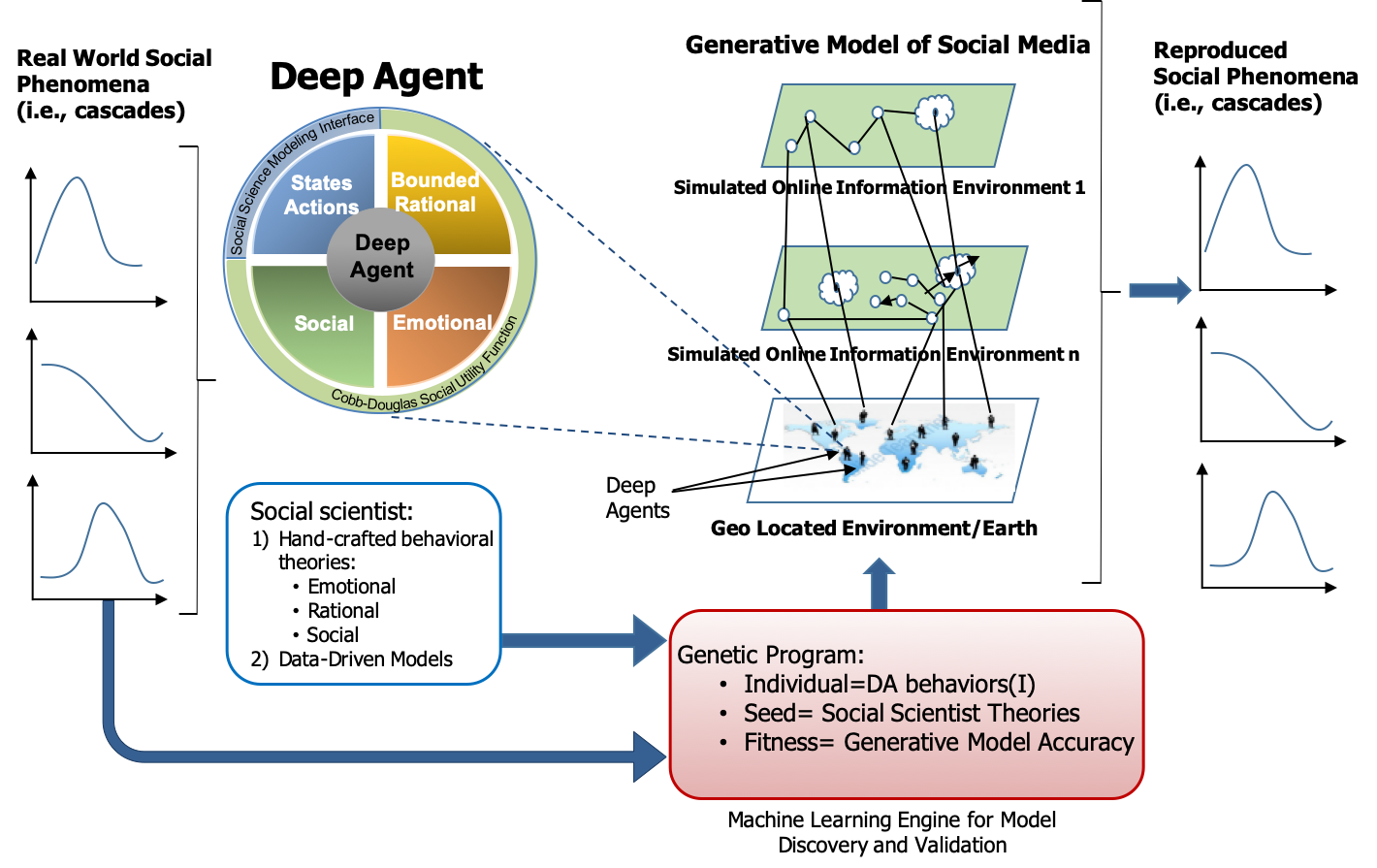}
            \caption[ ]
            {{\small}} 
            \label{fig:DAF_1(a)}
        \end{subfigure}
        \hfill
        \begin{subfigure}[b]{\textwidth}  
            \centering 
            \includegraphics[width=0.92\textwidth]{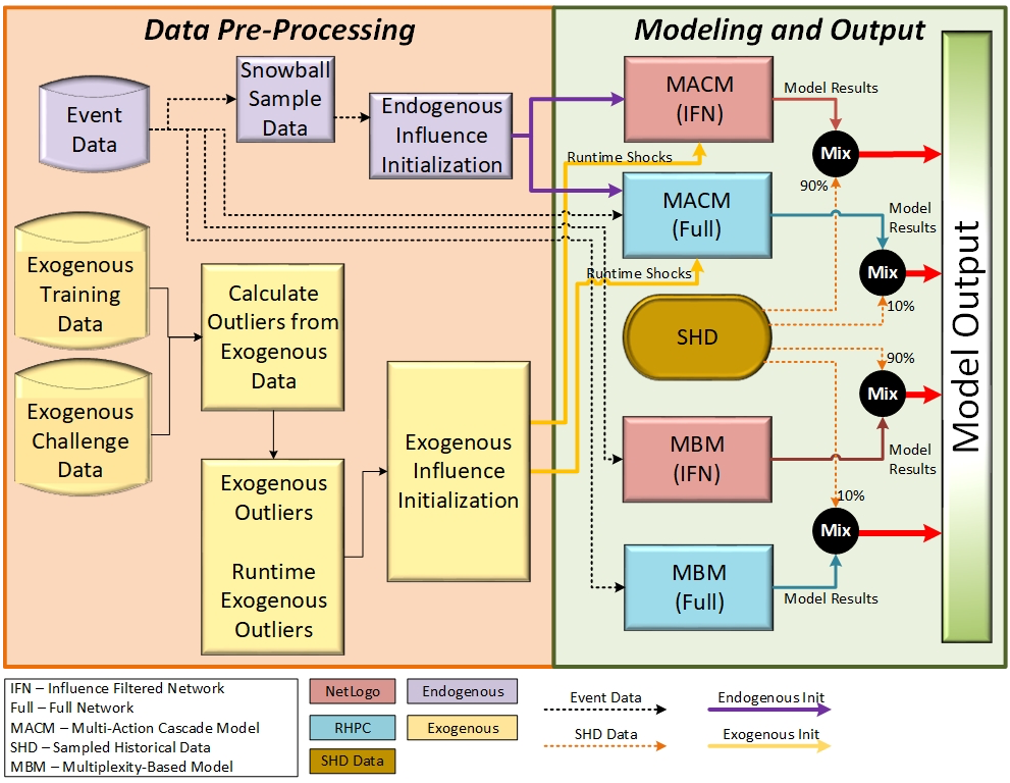}
            \caption[ ]
            {{\small}}    
            \label{fig:DAF_1(b)}
        \end{subfigure}
        \caption[  ]
        {\small Deep Agent Modeling Framework (DAF). (a): The framework maps the agents' rules of behavior to the real-world data using the hand-crafted behavioral theories (emotional, rational, and social) and the data-driven models to initialize the generative models. 
        \\(b): Architecture of the DAF
        is depicted, which consists of the work-flow of the data process (endogenous and exogenous data), model simulation, model mixing, model evaluation and tuning.} 
    \end{figure*}

Our generative model of social media (figure \ref{fig:DAF_1(a)}) employs ABMs to simulate social dynamics via embedded agents as user profiles in OSN platforms, and ``deep agents'' 
as the platform users. 
The deep agent concept is adopted from the Agent-Zero \cite{epstein2014agent_zero} and the Homos Socialis \cite{gintis2015homo} frameworks, which leads to deep agent modeling. Considering the Agent Zero framework, a cognitively plausible agent must account for three dimensions: Emotions (leading to ortho-rational behavior), Bounded Rationality, and Social Connectivity. Therefore, the social media users in our DAF framework were modeled to possess these three dimensions, and are referred as the deep agents in this paper. In contrast, ``shallow'' models are deficient in modeling deep human characteristics that would determine social behavior outcomes. 
Additionally, 
shallow 
models focus
on fitting equations 
into a single phenomenon of interest, which results in models that are brittle and potentially over-fitted for a single dimension of the data.

The agents' interactions according to the embedded rules result in specific outcomes at the population, community, user, and content levels, and provide information regarding the agents' decision processes. The agents' relative actions are derivable by applying particular metrics, (addressed in \ref{subsection:evaluation metrics}) and the Appendix sections, which provide simple statistics for the cascade and group behaviors.

The designed Deep Agent Modeling Framework (DAF), depicted in Figure \ref{fig:DAF_1(b)}, allows to create a family of modular sub-components from which multiple plausible models can be systematically assembled, tested, and validated. 
Discovering every rule of behavior is possible through employing the genetic programming evolutionary model discovery method (red box in Figure \ref{fig:DAF_1(a)}) as in \cite{gunaratne2017alternate}, which explores every possible space related to the set of agents' rules of behavior. This provides strong inferences of human behavior using computational simulations. 
The search can be guided by model accuracy, as measured by comparing model outputs with real world social dynamics data. Assembling the framework with the explained pieces, this framework unleashes the power of combining massively parallel computing, data analytics of large datasets and,
machine learning to assist in mixing and matching sub-models in a semi-automated way; this allows for exploration and testing that validates tens of thousands of models against a large set of target behaviors.

We developed the DAF simulation tool to help answering the questions related to properties of information exchange dynamics on online social media in population, content, user, and community levels. Additionally, we proposed the Multiplexity-Based Model (MBM), which captures social network evolution based on the preferential attachment, attention and recency cognitive bias. 
We mix and compare our model with other theory-driven models designed by our team of researchers. These models include Multi-Action Cascade Model (MACM) and Sampled Historical Data (SHD). The mixing, evaluation, and comparisons 
are provided for the three models and their variations: MBM-Influenced Filtered Network, MBM, SHD, MACM-Influenced Filtered Network, and finally MACM. 
In this paper we applied hand-crafted behavioral theories and data-driven models directly to feed 
the agent-based models.

 \subsection{Deep Agent Framework: Architecture and Analysis}
 The architecture of the framework (figure \ref{fig:DAF_1(b)}) consists of the data pre-processing, and the modeling and output phases. The data pre-processing phase includes data sampling, extracting influential users through normalized transfer entropy, extracting external shocks to the system represented as outliers, and initializing the endogenous and exogenous influences with the extracted users and shocks. 
The endogenous influence initialization involves the snowball sampling of the influential users and their relationships, to generate the static endogenous network, the network dynamics, and the network message information. 
Snowball sampling with normalized transfer entropy was used to extract the influential relationships from the event data, which was used beside the extracted activity disparity distributions of the endogenous relationships to build the static exogenous network. 
Using the extracted influential relationships, we used most recent activities to build the initialized network dynamics and the last $m$ messages to filter the network message information.

Each model has two variations, a full network simulation and an Influence Filtered Network (IFN) simulation. The full models take in the entire network of the event data in the initialization phase; however, the IFN models are initialized using only the filtered influential users to simulate the user interactions. 
We used all three outputs of the endogenous influence initialization phase as the inputs to the MACM and MACM-IFN models. 

For the exogenous data initialization process, we extracted the outliers from the exogenous training and challenge data sources using different filtering methods. This process includes applying Fourier transform (FT) on each different time-series, employing a moving window Magnitude filter and Butterworth filter on the FT of each different time-series to filter the anomalies, and the binary digitization of the anomaly time-series. We applied transfer entropy to the endogenous outliers and extracted the activity disparity distributions of the exogenous relationships from the filtered data to build the static exogenous network. Finally, we generated the network dynamics via extracting the last activity disparity of the exogenous relationships from the exogenous outliers. The generated static exogenous network and the network dynamics were fed
into MACM and MACM-IFN models.
The inputs to the MBM and MBM-IFN models were the entire network of the event data and the sampled data related to the last $x$ weeks, respectively.

The modeling and output phase contains our models' simulations and mixing of the models, model evaluation, and model tuning. The five variations of the generative models were implemented using Netlogo and RHPC coding environments. 
The model mixing strategy refers to merging the simulation outputs of the MACM and MBM models with the simulation output from the SHD model. 
The mixing strategy combines the output of the Full models with 10\% of the simulated user interactions using the SHD model. The IFN models take 90\% of the simulated interactions using the SHD model as input combined with the model results. More information on mixing strategy is provided in subsection \ref{shd}.

\subsection{Agent-Based Models}
The agent-based models in this paper are generative rule-driven models, designed based on the social theory on Diffusion of Information (DoI) and the user actions in OSNs. Although each social media platform has specific user interactions, we can refer to four fundamental user activities observable
in any OSN platform: Create, Post, Vote, and Follow.
Along with this, there are four entities in any OSN environment: Actor, Content, Action, and Space. This viewpoint of the actions and entities allows behavior to be represented
across multiple social media environments, referred to as the common language or the ontology of user actions and entities. The agent-based models in this paper are designed based on the four traditional DoI models: 
the Threshold Model (TM) \cite{granovetter1978threshold, watts2002simple, bass2004new};
the Independent Cascade Model (ICM) \cite{goldenberg2001using, goldenberg2001talk},
the Bass Growth Model \cite{bass1969new, bass2004new}, Rand Agent-Based Model \cite{rand2011agent}, and the Complex Contagion Model \cite{centola2007complex}. 

\subsubsection{Multiplexity-Based Model}
\begin{table}[t]
  \caption{Table of symbols and definitions used in this paper.}
  \centering
  \begin{tabular}{ p{6cm} c}
    \toprule
    Variable definition & Symbol \\
    \midrule
    Activity/Action           &    $A$ \\
    Message Content       &    $M$ \\
    Conversation              &    $C$ \\
    External Shock            &    $S$ \\
    Fitness                   &    $F$\\
    Link/Edge                 &    $L$\\

    Node Degree               &    $K$\\
    
    Stochastic Noise          &    $\epsilon$ \\
    Time-step/Tick            &    $t$\\
    Transfer Entropy          &    $T$ \\

    User/Node                  &    $U$ \\ 
    Vote to a Content &   $V$ \\


    \bottomrule
  \end{tabular}
\end{table}

The MBM model simulates social network evolution by multiplex networks, which have multi-layer network structure with possible shared nodes among different layers \cite{basu2015multiplex}.
As MBM is designed based on concepts from graph theory, we refer to OSN users as nodes and user interactions as links.  The model consists of a directed bipartite graph with bipartite pairs of users-repositories for GitHub, users-subReddits for Reddit, and users-users for Twitter, distinguished by multiple layers. Each of the separate user actions in the platform generate a sub-graph and the combination of the actions generate the whole network structure.
The set of user actions in this model are 
conversation creation, contribution, vote
, and follow,   
which can be formalized as $(C_i \notin \{C\})$, $(C_i = C_j , M_i \notin \{M\})$,
$(V_i \notin \{V\})$, and $(L_i \notin \{L_{U_j}\})$
respectively, where indices are representative of users that perform the action, and $\{C\}$, $\{V\}$, and $\{L_{U_j}\}$ refer to the sets of all conversations, votes, and links to followers of the user in the model up to the current time-step. 

The cognitive factor of MBM refers to the information overload resulting in higher attention to recent activities and active users. In other words, MBM considers the recency bias affecting OSN users' decision making processes to possible propagation of information. This concept has been designed in the model in terms of age and fitness values, such that the user's influence decays in time.
Content targets that have been recently the object of actions, and the users that have recently acted, see their fitness decrease the least, whereas these values for inactive users are decreased the most. This results in paying higher attention to the influential users and targets, but allowing their fitness values to be reduced in popularity over time, and eventually to be supplanted by newer elements. 
Reaching a certain age leads to the node removal from the model node-set. As a result, the model's predictions are most affected by recent trending activities with higher attention to more active users. 
Accordingly, the driving forces of MBM are preferential attachment and preferential decay, both as functions of the node's degree $k \in \{1, \dots, K\}$ and age $a \in \{t_0, \dots, t_{max}\}$.
The propagation of information from user $U_j$ to neighboring user $U_i$ can be represented as:


\begin{equation}
(U_j, A_j, a_j) \rightarrow (U_i, A_i, a_i)
\end{equation}

The model initially comprises of 
$|U|$
nodes, 
with each node as a user $U$ having 
$L_U$ 
number of links.
Each node in the graph is assigned a string fitness of $F = 1$ that models the node's influence on the growth of the network and decreases as a function of time, node age, and activity history \cite{mcpherson2001birds}. Accordingly, node's age value can be calculated as:
\begin{equation}
a_j \leftarrow a_j + \Big(1- (t_{c_j} - t_{p_j}) * F_j)\Big)
\end{equation} 
where $t_{c_j}$ is the current time and $t_{p_j}$ refers to the previous activity time for user $U_j$, and $F_j$ is user's current fitness value. 
The value of fitness for each node can be calculated as:
\begin{equation}
 F_j = \frac{|A_j|}{a_j}   
\end{equation}

where $|A_j|$ is the number of actions for user $U_j$ and is equal to the number of user interactions or degree $k_j$.  
In other words, fitness is essentially a simulation of a user's productivity. User’s fitness is increased by the diversity of activities, the shortness of time-span between its activities, and the fitness of all the interactions a user builds.


MBM network grows at each time-step by the successive addition of new nodes to the model node-set and new edges to the edge-set. Node addition ratio 
was extracted from the input data. 
New links emerge between the nodes with higher fitness values as a result of the preferential attachment.


In summary, the model can be broken down into three steps that are performed at each time-step: 
 \begin{enumerate*}[label={\Roman*)}]
\item Node selection: selecting a set of nodes from the set of all current model node-set and potential nodes that can be added to this set, referred here as 
$\{U_t\} \subset \{U\}$. 

\item Interaction: building the interactions between the bipartite node pairs in $\{U_t\}$ and the rest of nodes in the model such that the pairs are matched according to a likelihood distribution weighted on the nodes' fitness values. A sub-graph associated with a behavior activity is assigned to the selected pair based on a likelihood distribution determined by the popularity of action types. The layer fitness score for each sub-graph gets re-calculated to predict how popular each action remains.;

\item Update: 
updating the node and link fitness scores, local and global degree values, layer fitness scores, node and link ages, and removing the nodes and edges from the model according to fitness decay. In this step, the new age value for each node can be calculated by:
\end{enumerate*}
\begin{equation}
a_j \leftarrow a_j + \Big(1- (t_{c_j}-t_{p_j}) * (t_{c_j} + 1)\Big)
\end{equation}

\subsubsection{Multi-Action Cascade Model}
The MACM model \cite{gunaratne2019multi, gunaratne2019theory} is a cognition-based agent-based model that simulates the diffusion of information through the network using individual-scale probabilities of actions derived from the Independent Cascade Model. The cognitive factor of this model refers to the information overload resulted by vast amount of social media activity bombarding users' attention and affecting their decision making processes through prioritizing and preference to possible propagation of information. 

Using transfer entropy analysis on user-user and user-exogenous force influences,  this model measures the probabilities of actions for user $U_i$ related to endogenous and exogenous forces as:
\begin{equation}
q = \mathbb{P}_t(U_i | U_j) = \mathbb{P}_{t - 1}(U_i | U_j) + \frac{\epsilon_{i, j}}{1 + T_{i, j}}\; 
\end{equation}

\begin{equation}
p = \mathbb{P}_t(U_i | S) = \underset{s \in S}{\bigcup} \Big( \mathbb{P}_{t-1}(U_i | s) + \frac{\epsilon_{i, s}}{1 + T_{i, s}} \Big)
\end{equation}

where neighboring user $U_j$ is active in a conversation, $s \in S$ refers to external shock, $T_{i, j}$ is the transfer entropy from user $U_j$'s action to user $U_i$'s action, $T_{i, s}$ is the transfer entropy from external shock $s$ to user $U_i$'s action, and $\epsilon$ indicates noise relative to activity changes of the two users.

MACM considers the evidence that the internal and the external forces can define the dynamics of different event types causing the spread of information, and the evolution of a content as it spreads through the social network.
The user actions in this model are conversation creation, contribution, sharing, and deletion, 
which can be formalized as $(C_i \notin \{C\})$, $(C_i = C_j , M_i \notin \{M\})$, $(C_i != C_j , M_i = M_j)$, and $(C_i = C_j , M_i = \o)$ respectively, where indices are representative of users that perform the action, and $\{C\}$ and $\{M\}$ refer to the sets of all conversations and contents in the model up to the current time-step. 
The propagation of information is modeled as a message considering the influences from the neighboring nodes, the action type, the target conversation, and the content genome, such that: 
\begin{equation}
(U_j, A_j, C_j, M_j) \rightarrow (U_i, A_i, C_i, M_i)
\end{equation}
where $Cont_A$ is the massage content in conversation $C$, $A$ refers to the action, and $U$ represents the user. Additionally, the user interactions are conditioned on the processing of the received messages from the connected nodes filtered down as a result of the cognitive overloading.
Accordingly, MACM agents calculate the probability of performing action $A_i$ in response to action $A_j$ as the union of probabilities of actions based on the endogenous and exogenous forces as:
\begin{equation}
    \mathbb{P}_t(U_i | U_j, S) = \mathbb{P}_t(U_i | U_j) \cup \mathbb{P}_t(U_i | S)
\end{equation}\;

\subsubsection{Sampled Historical Data}
\label{shd}
The SHD model \cite{NedaSHD, saadat2018initializing} is a replay-based data mixture model designed based on the seasonality 
characteristic of the OSN user activities and the hypothesis that the users exhibit repetitive patterns. This model extracts the most recent activities from the training data to provide the information related to the user interactions and edge formations in the network, and predicts the future user interactions according to the same types of activities in the past. 
We employed the SHD model to simulate the less-active users that hold little influence on the network and have been removed in the filtering processes.    
The mixing strategy using the SHD model refers to:
\begin{enumerate*}[label={\Roman*)}]
    \item extracting the active and less-active unique users from the social network data; 
    \item predict the activities associated with the active users using MBM and MACM models; 
    \item predict the activities associated with the less-active users using SHD; and 
    \item append the SHD simulated low-activity users to the events simulated by the MBM and MACM models.
\end{enumerate*}

\section{Dataset Description}
The challenge goal was to model social structures and their day to day changes, and accordingly, simulate the time-series network evolution of GitHub, Twitter and Reddit social environments for the three domains of interest (CVEs, Cyber Threats, and Cryptocurrencies).
The datasets for the challenge were provided by Leidos, and are explained in detail below.

The GitHub social network data contained information from the
years 2015 to 2017.  A total of 33,570 cryptocurrency-related repositories were associated with or included target coin names or keywords in their descriptions, and 1,193,370 events matched with those repositories.
5,505,496 cybersecurity repositories and 214,074,771 events were selected as well as 186,190 software vulnerability related repositories and 26,777,997 events.
The Twitter social network included data for the years 2016 to 2017, and with a total of 7,382,724 cryptocurrency-related tweets, retweets, and quotes.
These values were 30,704,025 and 74,074 for the cybersecurity and software vulnerability domains respectively. 
The Reddit dataset included submissions and comments matching keywords for the years 2015 to 2017. 
The cryptocurrency-related data contained 299,401 submissions and 3,370,547 comments. These values were 2,442,942 and 33,629,588 for the cybersecurity domain, and 60,760 and 264,024 for the software vulnerability domain. 

\subsection{Evaluation Events and Metrics}
Since each social environment has a unique set of events depending on the interactions in the platform, the evaluation events for each social environment were defined separately as the following:
\begin{enumerate*}[label={\Roman*)}]
\item GitHub: Commit Comment, Create, Delete, Fork, Issue Comment, Push, Watch, Pull Request, and Issues, 
\item Twitter: Tweet (create original material),  Retweet, Reply, and Quote,  
\item Reddit: Comment and Post.
\end{enumerate*}

\label{subsection:evaluation metrics}
The evaluation measurements were applied to online social behaviors at the population, content, user, and community levels.
Content-based measurements are any user interaction, that is,
posting or replying in Reddit and Twitter platforms, or writing a comment in Github. User-level measurements were focused on user activities; for instance, the contribution counts of the user over time. Finally, the population-level measurements aggregated the events' and users' characteristics on a particular platform. 
Examples of the accuracy measurements and metrics include community burstiness and user Gini coefficient calculated by absolute percentage error, community Gini measured by absolute difference, user trustingness, user diffusion delay calculated by the Kolmogorov-Smirnov test (K-S test), and user popularity measured by ranked-biased overlap (RBO).
Additionally, ``surprise measurements'' were provided during the test event for competing campaigns in the cryptocurrency domain and competing attention in the cyber threats and CVE domains.

\section{Experimental Results}
\begin{figure}
    \centering
    \includegraphics[width=0.8\linewidth]{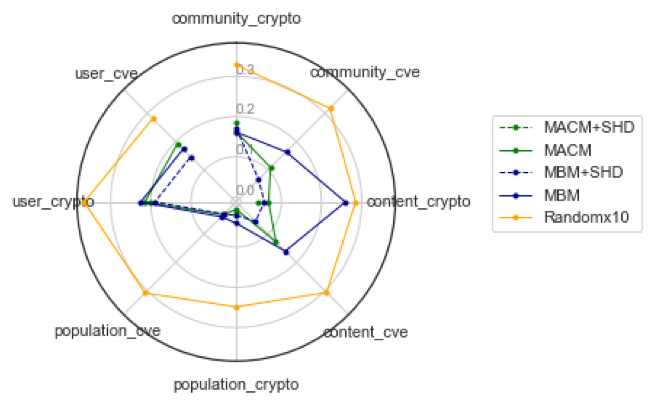} 
    \caption[ ]
    {{\small 
    The comparison of the normalized error ratio for the agent-based models and the ground truth based on the measurement types. A lower value corresponds to better model performance.
    
    }}
    \label{fig:overal_benchmark}
\end{figure}

\begin{figure}
\centering
\begin{subfigure}[b]{\textwidth}
    \centering
    \includegraphics[width=0.7\linewidth]{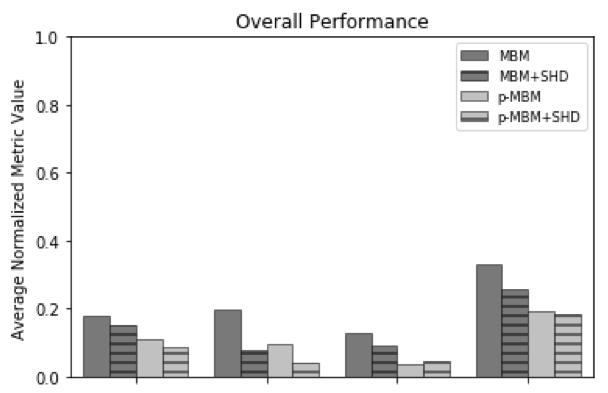} 
    \caption[ ]
    {{\small}}
    \label{fig:overal_p_a}
\end{subfigure}
\begin{subfigure}[b]{\textwidth}
    \centering
    \includegraphics[width=0.7\linewidth]{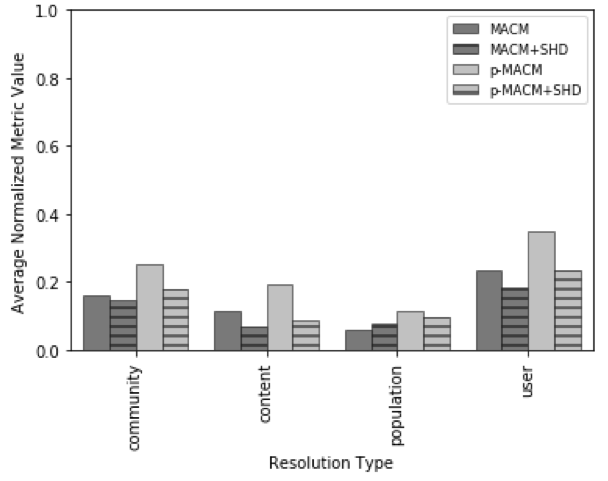} 
    \caption[ ]
    {{\small}}
    \label{fig:overal_p_b}
\end{subfigure}
\caption[]{{\small Overall performance comparison over the (a) MBM and (b) MACM model variations over the community, content, population, and user levels, depicted in terms of normalized error ratio. A lower value corresponds to better model performance.}}
\end{figure}

We applied extensive analysis using our framework to mix and match and compare the models. We also calculated model performances for the user, community, population, and content levels for the cybersecurity, software vulnerability, and cryptocurrency domains. 
The training data input for the agent-based models were GitHub, Twitter and Reddit network evolution over time for the three domains of interest, and the exogenous data for both training and test periods. 
Simulation inputs include the initialization data, the exogenous data, and the events data.  
There were two kinds of expected simulation outputs: 
\begin{enumerate*}[label={\Roman*)}]
\item full network dynamics (event logs/data frames obeying specific formats that contain event-related information), and \item direct output of the accuracy measurements. 
\end{enumerate*}

The corresponding results for the benchmark comparisons are provided in Figure \ref{fig:overal_benchmark}, which represents the comparison of the model measurements versus the ground truth (Random) considering the measurement types. The Jensen-Shannon Divergence, Kolmogorov-Smirnov test and absolute percentage error metrics were among the metrics used for this analysis. 
We normalized the model output results over the measurement group (user, community, population, and content), metric type, and the platform (GitHub, Twitter, Reddit) to calculate a value in the range 0 to 1, where the lower value corresponds to better model performance.

Figures \ref{fig:overal_p_a}- \ref{fig:overal_p_b} demonstrate the model-wise performance comparisons with and without the SHD strategy mixing, and for the two user influence pruning cases. The prefix ``p'' refers to the models initialized using the data pruned for the influential users, while the non-prefixed models relate to the models initialized employing the entire event data. We can observe that pruning influential relationships improved the MBM model performance; however, this strategy was not successful with the MACM model. 
The results indicate that social theory-based modeling may consider influential interactions inherently, and support that the more ``explainable'' a model, the higher the performance. Additionally, the results provide evidence that the mixing strategy helps improve the user and community level performances, and the single models are more successful in modeling the population and content level interactions of the influential users.

The metric-based performance comparison of the models are provided in table \ref{tab:1} in the Appendix section, for the community, content, population, and user level interactions, respectively. In these figures, the rows refer to the group-based measurement metrics for the models in rows. Each occupied cell indicates the best performance of the relative model for the specific metric. The values refer to the normalized sub-metrics averaged over 105 model runs. The content and user level performances illustrate that the mixing strategy using the Standard Historical Data (SHD) improves the model performance in modeling the node level interactions. Finally, the population level scores illustrate another benefit of the SHD mixing strategy in improving the performance for the degree distributions and the node level characteristics explanations.

\section{Conclusion and Future Work}
In this paper, we discussed how user interactions, behaviors, and complex human dynamics can be captured via combining massively parallel computing, data analytics of large datasets, and machine learning algorithms. We proposed the Deep Agent Framework (DAF), which operates beyond single models by mixing and matching sub-models in a semi-automated way. Our framework operationalizes social theories of human behavior and social media into optimized generative simulation capabilities that enable exploring information diffusion and evolution within the social media context. 
Our multi-resolution simulation at the user, community, population, and content levels, and our extensive analysis and results provide evidence that our framework is a powerful tool in modeling the diffusion and evolution of the information in variety of online social platforms. Although, we applied hand-crafted behavioral theories and data-driven models directly to feed the agent-based models without employing the genetic program (red box in Figure \ref{fig:DAF_1(a)}), our results prove that our framework and our deep generative models are powerful in modeling online social network interactions. 

Further improvements to the DAF framework can be made by employing the evolutionary model discovery to explore the space for the rule sets of behaviors related to the agents, which allows for the testing and validation of tens of thousands of models against large set of target behaviors as in \cite{gunaratne2017alternate}.
Additionally, the future direction of our
work serves to automatically introduce variants to all the models produced by different performers to obtain the overall best model.

\begin{acknowledgement}
This work was supported by the Defense Advanced Research Projects Agency (DARPA) under grant number FA8650-18-C-7823. The views and opinions expressed in this article are the authors' own and should not be construed as official or as reflecting the views of the University of Central Florida, DARPA, or the U.S.
Department of Defense.
\end{acknowledgement}

\section*{Appendix}
\label{apendix}
\addcontentsline{toc}{section}{Appendix}
This section provides a list of measurement metrics across community, content, population, and user levels in table \ref{tab:1}. Each occupied cell indicates the relative performance of the model in terms of error (in columns) reported for a specified metric (in rows). Values in the cells refer to the error ratio assessed via first normalizing over each sub-metric, followed by averaging over the results for the model runs. In this table, a lower value corresponds to better model performance and the error ratios below 0.2 are marked in bold.

\begin{table}
\caption{Performance comparison for the Agent-based models across Community, Content, Population, and User level metrics.}
\label{tab:1}       
%
%
\begin{tabular}{p{1.1cm}|p{4.2cm}|p{1cm}p{1.7cm}p{1.2cm}p{2cm}}
\hline\noalign{\smallskip}
Category & Metric & MBM & MBM + SHD & pMACM & pMACM + SHD  \\
\noalign{\smallskip}\svhline\noalign{\smallskip}
\multirow{10}{*}{\rotatebox[origin=c]{90}{Community}} 
& Burstiness & 0.29  & 0.22 & 0.34 & 0.26 \\
& Contributing users & 0.31  & 0.24 & 0.35 & 0.26\\
& Event proportions & 0.38  & \textbf{0.20} & 0.38 & 0.26\\
& Geo locations & 0.22  & 0.23 & 0.33 & \textbf{0.17}\\
& Gini coefficient &  0.26 & \textbf{0.18} & 0.29 & \textbf{0.20}\\
& Issue types &  \textbf{0.00} & 0.30 & \textbf{0.00} & 0.27\\
& User action counts & 0.37 & 0.25  & 0.50 & 0.31 \\
& Palma coefficient  & 0.55  & 0.31 & 0.53 & 0.40 \\
& User account ages  &  \textbf{0.14} & \textbf{0.14} & 0.54 & \textbf{0.15} \\
& User burstiness  & \textbf{0.18}  & 0.36 & 0.65 & 0.31 \\
\noalign{\smallskip}\svhline\noalign{\smallskip}
\multirow{13}{*}{\rotatebox[origin=c]{90}{Content}} 
& Activity disparity Gini coefficient & \textbf{0.06}  & \textbf{0.07} & \textbf{0.04} & \textbf{0.03} \\
& Activity disparity Palma coefficient & 0.28  & \textbf{0.12} & \textbf{0.09} & \textbf{0.16}\\
& Contributors & \textbf{0.06}  & \textbf{0.10} & \textbf{0.11} & \textbf{0.00}\\
& Diffusion delay & 0.22  & \textbf{0.11} & 0.63 & \textbf{0.11}\\
& Event counts &  0.29 & \textbf{0.05} & \textbf{0.01} & \textbf{0.00}\\
& Daily event distribution &  0.85 & 0.31 & 0.80 & 0.30\\
& Day of Week event distribution  & 0.78  & 0.22 & 0.75 & 0.20\\
& Growth  & \textbf{0.07}  & \textbf{0.12} & \textbf{0.16} & \textbf{0.01}\\
& Liveliness distribution  & \textbf{0.15}  & \textbf{0.06} & \textbf{0.15} & \textbf{0.01}\\
& Liveliness top K  &  0.90 & 0.52 & 0.86 & 0.38\\
& Popularity distribution top K  &  \textbf{0.10} & \textbf{0.03} & 0.26 & \textbf{0.06}\\
& Popularity top K  &  0.78 & 0.55 & 0.89 & 0.64\\
& User unique content  & \textbf{0.06}  & \textbf{0.07} & \textbf{0.10} & \textbf{0.05}\\
\noalign{\smallskip}\svhline\noalign{\smallskip}
\multirow{11}{*}{\rotatebox[origin=c]{90}{Population}} 
& Assortativity coefficient & \textbf{0.03}  & \textbf{0.01} & \textbf{0.01} & \textbf{0.01} \\
& Average clustering coefficient & \textbf{0.01}  & \textbf{0.02} & \textbf{0.01} & \textbf{0.01}\\
& Community modularity & \textbf{0.01}  & \textbf{0.01} & \textbf{0.01} & \textbf{0.01}\\
& Degree distribution & 0.81  & 0.68 & 0.34 & 0.54\\
& Density &  \textbf{0.02} & \textbf{0.01} & \textbf{0.11} & \textbf{0.01}\\
& Max node degree &  \textbf{0.03} & \textbf{0.02} & \textbf{0.04} & \textbf{0.03}\\
& Mean node degree  &  \textbf{0.01} & \textbf{0.01} & \textbf{0.02} & \textbf{0.01}\\
& Mean shortest path length  & \textbf{0.01}  & \textbf{0.01} & \textbf{0.00} & \textbf{0.01}\\
& Number of connected components  & \textbf{0.06}  & \textbf{0.10} & \textbf{0.01} & \textbf{0.02}\\
& Number of edges  & \textbf{0.15}  & \textbf{0.13} & \textbf{0.14} & \textbf{0.13}\\
& Number of nodes  &  \textbf{0.07} & \textbf{0.08} & \textbf{0.03} & \textbf{0.02}\\
\noalign{\smallskip}\svhline\noalign{\smallskip}
\multirow{11}{*}{\rotatebox[origin=c]{90}{User}} & 
Most active users & 0.98  & 0.76 & 0.97 & 0.71 \\
& Repository user continue proportion & \textbf{0.06}  & \textbf{0.01} & \textbf{0.03} & \textbf{0.01}\\
& Subreddit user continue proportion & \textbf{0.02}  & \textbf{0.01} & \textbf{0.00} & \textbf{0.00}\\
& Activity distribution & 0.31  & 0.32 & 0.44 & 0.32\\
& Activity timeline &  0.30 & \textbf{0.18} & 0.38 & \textbf{0.12}\\
& Diffusion delay &  0.47 & \textbf{0.14} & 0.35 & 0.46\\
& Gini Coefficient  &  \textbf{0.10} & \textbf{0.07} & \textbf{0.03} & \textbf{0.01}\\
& Palma Coefficient  &  0.26 & 0.26 & 0.29 & 0.30 \\
& Popularity  & 0.82  & 0.69 & 0.85 & 0.67 \\
& Trustingness  & \textbf{0.00}  & \textbf{0.08} & \textbf{0.00} & \textbf{0.08}\\
& User unique content  & \textbf{0.05}  & \textbf{0.07} & \textbf{0.19} & \textbf{0.07}\\
\noalign{\smallskip}\hline\noalign{\smallskip}
\end{tabular}
\end{table}
\footnotesize {
%
%

\begin{thebibliography}{99.}%

\bibitem{adamic2016information}
Lada~A Adamic, Thomas~M Lento, Eytan Adar, and Pauline~C Ng.
\newblock Information evolution in social networks.
\newblock In {\em Proceedings of the Ninth ACM International Conference on Web
  Search and Data Mining}, pages 473--482. ACM, 2016.


\bibitem{bass1969new}
Frank~M Bass.
\newblock A new product growth for model consumer durables.
\newblock {\em Management science}, 15(5):215--227, 1969.

\bibitem{bass2004new}
Frank~M Bass.
\newblock A new product growth for model consumer durables.
\newblock {\em Management science}, 50(12\_supplement):1825--1832, 2004.

\bibitem{centola2007complex}
Damon Centola and Michael Macy.
\newblock Complex contagions and the weakness of long ties.
\newblock {\em American journal of Sociology}, 113(3):702--734, 2007.


\bibitem{epstein2014agent_zero}
Joshua~M Epstein.
\newblock {\em Agent\_zero: Toward neurocognitive foundations for generative
  social science}, volume~25.
\newblock Princeton University Press, 2014.


\bibitem{gintis2015homo}
Herbert Gintis, Dirk Helbing, et~al.
\newblock Homo socialis: An analytical core for sociological theory.
\newblock {\em Review of Behavioral Economics}, 2(1-2):1--59, 2015.

\bibitem{goldenberg2001talk}
Jacob Goldenberg, Barak Libai, and Eitan Muller.
\newblock Talk of the network: A complex systems look at the underlying process
  of word-of-mouth.
\newblock {\em Marketing letters}, 12(3):211--223, 2001.

\bibitem{goldenberg2001using}
Jacob Goldenberg, Barak Libai, and Eitan Muller.
\newblock Using complex systems analysis to advance marketing theory
  development: Modeling heterogeneity effects on new product growth through
  stochastic cellular automata.
\newblock {\em Academy of Marketing Science Review}, 9(3):1--18, 2001.

\bibitem{granovetter1978threshold}
Mark Granovetter.
\newblock Threshold models of collective behavior.
\newblock {\em American journal of sociology}, 83(6):1420--1443, 1978.

\bibitem{gunaratne2017alternate}
Chathika Gunaratne and Ivan Garibay.
\newblock Alternate social theory discovery using genetic programming: towards
  better understanding the artificial anasazi.
\newblock In {\em Proceedings of the Genetic and Evolutionary Computation
  Conference}, pages 115--122. ACM, 2017.

\bibitem{gunaratne2019multi}
Chathika Gunaratne, Chathurani Senevirathna, Chathura Jayalath, Nisha Baral,
  William Rand, and Ivan Garibay.
\newblock A multi-action cascade model of conversation.
\newblock In {\em 5th International Conference on Computational Social Science,
  URL http://app. ic2s2. org/app/sessions/9kXqn5btgKKC5yfCvg/details}, 2019.

\bibitem{NedaSHD}
Neda Hajiakhoond~Bidoki, Madeline Schiappa, Gita Sukthankar, and Ivan Garibay.
\newblock Predicting social network evolution from community data partitions.
\newblock {\em 2019 International Conference on Social Computing,
  Behavioral-Cultural Modeling and Prediction and Behavior Representation in
  Modeling and Simulation}, pages In--press, 2019.


\bibitem{rand2015agent}
William Rand, Jeffrey Herrmann, Brandon Schein, and Ne{\v{z}}a Vodopivec.
\newblock An agent-based model of urgent diffusion in social media.
\newblock {\em Journal of Artificial Societies and Social Simulation}, 18(2):1,
  2015.

\bibitem{rand2011agent}
William Rand and Roland~T Rust.
\newblock Agent-based modeling in marketing: Guidelines for rigor.
\newblock {\em International Journal of Research in Marketing}, 28(3):181--193,
  2011.


\bibitem{vsuvakov2012agent}
Milovan {\v{S}}uvakov, David Garcia, Frank Schweitzer, and Bosiljka Tadi{\'c}.
\newblock Agent-based simulations of emotion spreading in online social
  networks.
\newblock {\em arXiv preprint arXiv:1205.6278}, 2012.

\bibitem{watts2002simple}
Duncan~J Watts.
\newblock A simple model of global cascades on random networks.
\newblock {\em Proceedings of the National Academy of Sciences},
  99(9):5766--5771, 2002.


\bibitem{basu2015multiplex}
Prithwish Basu, Matthew Dippel, and Ravi Sundaram.
\newblock Multiplex networks: A generative model and algorithmic complexity.
\newblock In {\em 2015 IEEE/ACM International Conference on Advances in Social
  Networks Analysis and Mining (ASONAM)}, pages 456--463. IEEE, 2015.

\bibitem{gunaratne2019theory}
Chathika Gunaratne, Nisha Baral, William Rand, Ivan Garibay, Chathura Jayalath,
  and Chathurani Senevirathna.
\newblock The effects of information overload on online conversation dynamics.
\newblock {\em Computational and Mathematical Organization Theory}, 2020.

\bibitem{mcpherson2001birds}
Miller McPherson, Lynn Smith-Lovin, and James~M Cook.
\newblock Birds of a feather: Homophily in social networks.
\newblock {\em Annual review of sociology}, 27(1):415--444, 2001.

\bibitem{saadat2018initializing}
Samaneh Saadat, Chathika Gunaratne, Nisha Baral, Gita Sukthankar, and Ivan
  Garibay.
\newblock Initializing agent-based models with clustering archetypes.
\newblock In {\em International Conference on Social Computing,
  Behavioral-Cultural Modeling and Prediction and Behavior Representation in
  Modeling and Simulation}, pages 233--239. Springer, 2018.





\end{thebibliography}
%

}

\end{document}